\documentclass[aps,prb,superscriptaddress,twocolumn]{revtex4}

\usepackage{amsmath}
\usepackage[pdftex]{graphicx}
\usepackage{epstopdf}
\usepackage{xcolor}
\usepackage{booktabs}
\usepackage[abs]{overpic}

\def \bn{\begin{align}}
\def \en{\end{align}}
\def \be{\begin{equation}}
\def \ee{\end{equation}}
\def \bea{\begin{eqnarray}}
\def \eea{\end{eqnarray}}
\def \ba{\begin{array}}
\def \ea{\end{array}}

\def \av#1{{\langle#1\rangle}}

\def \ket#1{{|#1\rangle}}

\def \a{{\alpha}}

\def \yd{^\dagger}

\def \nn{{\nonumber}}

\usepackage{ulem}

\renewcommand{\epsilon}{\varepsilon}

\begin{document}
\title{Dicke phase transition without total spin conservation}
\author{Emanuele G. Dalla Torre}
\affiliation{Department of Physics, Bar Ilan University, Ramat Gan 5290002, Israel}
\author{Yulia Shchadilova}
\affiliation{Department of Physics, Harvard University, Cambridge, MA 02138, U.S.A.}
\author{Eli Y. Wilner }
\affiliation{Department of Physics, Columbia University, New York, NY 10027, U.S.A.}
\author{Mikhail D. Lukin}
\affiliation{Department of Physics, Harvard University, Cambridge, MA 02138, U.S.A.}
\author{Eugene Demler }
\affiliation{Department of Physics, Harvard University, Cambridge, MA 02138, U.S.A.}

\begin{abstract}


We develop a new fermionic path-integral formalism to analyze the phase diagram of open nonequilibrium  systems. The formalism is applied to analyze an ensemble of two-level atoms interacting with a single-mode optical cavity, described by the Dicke model. While this model is often used as the paradigmatic example of a phase transition in driven-dissipative systems, earlier theoretical studies were limited to the special case when the total spin of the atomic ensemble is conserved. This assumption is not justified in most experimental realizations. Our new approach allows us to analyze the problem in a more general case, including the experimentally relevant case of dissipative processes that act on each atom individually and do not conserve the total spin. We obtain a general expression for the position of the transition, which contains as special cases the two previously known regimes: i) non-equilibrium systems with losses and conserved spin and ii) closed systems in thermal equilibrium and with the Gibbs ensemble averaging over the values of the total spin. We perform a detailed study of different types of baths and point out the possibility of a surprising  non-monotonous dependence of the transition on the baths' parameters.

\end{abstract}
\date{\today}
\maketitle

{\bf Introduction} Understanding phase transitions in open quantum systems is a challenging problem at the interface of quantum optics, condensed matter, and atomic physics. In contrast to equilibrium phase transitions, which have been well understood using  powerful theoretical tools such as renormalization group approaches and conformal field theories, we still lack reliable theoretical tools for analyzing non-equilibrium open systems.  This makes it particularly important to analyze systems with known experimental realizations that allow direct comparison between theoretical predictions and experimental measurements. Two important examples of such systems are the directed percolation and the driven dissipative Dicke
model, which have been respectively realized in liquid crystals \cite{takeuchi2007directed,henkel2008non} and
quantum optics\cite{black03,baumann10,baumann2011exploring,brennecke2013real,baden2014realization,klinder2015dynamical,klinder2015observation}. In the case of the Dicke model, theoretical approaches that have been developed so far rely on the existence of an integral of motion, the total angular momentum, which significantly reduces the complexity of the problem\cite{hepp73,wang73,narducci1973energy,charmichael1973higher,duncan1974effect,hillery1985semiclassical,emary03}. In contrast, actual experiments involve dissipative processes that do not respect this conservation law, such as dissipative baths coupled to each individual atom. Their description requires more advanced theoretical tools.

\begin{figure}[b]
\includegraphics[scale=0.27]{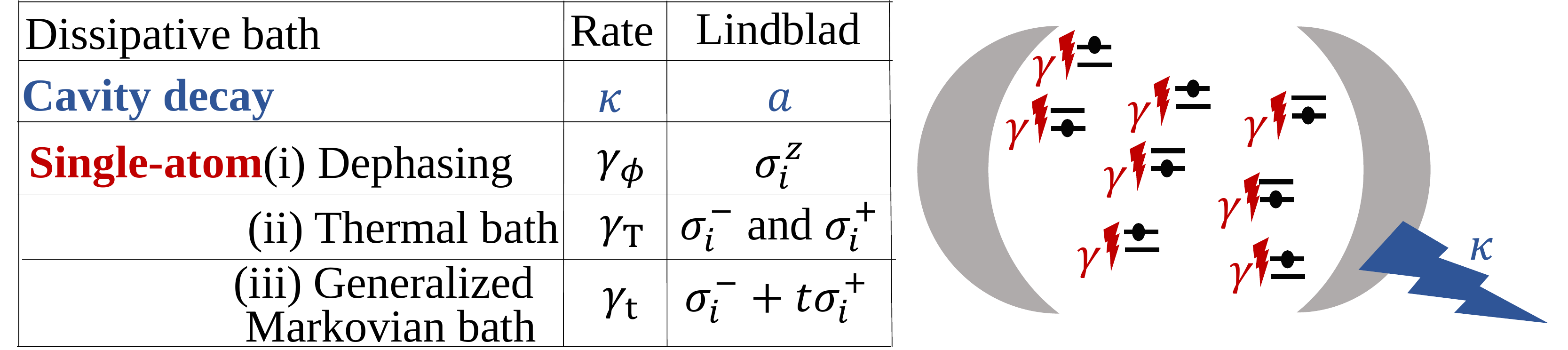}
\caption{Dissipative processes considered in the present study. Previous studies mostly focused on the cavity decay, which conserves the total spin.}
\label{fig:schematic}
\end{figure}

The effects of single-atom baths on the Dicke model were first considered in Refs.~[\onlinecite{dalladiehl},\onlinecite{strack2016}], using approximate methods based on effective {\it bosonic} field theories. These approaches map the two-level systems to continuous variables and are valid only if all the atoms are strongly polarized in a given direction\cite{strack2016}. In this paper we instead employ an exact mapping to a {\it fermionic} path-integral representation, which allows us to obtain an exact expression for the location of the Dicke transition. In the limit of a large number of atoms we recast our result in terms of single-atom correlation functions, which can be computed using standard master equations. The present approach reproduces the known position of the equilibrium phase transition and additionally allows us to systematically describe single-atom dephasing and decay (see Fig.~\ref{fig:schematic}). As we will show, these processes renormalize the position of the Dicke transition and in some cases completely destroy it.

{\bf Model} The Dicke model describes the interaction of $N$ two-level atoms (or spins), $\sigma_j$, with a single bosonic degree of freedom, $a$, 

\be H=\omega_0a\yd a+\omega_z\sum_{j=1,N}\sigma_j^z+\frac{2g}{\sqrt{N}}\sum_{j=1}^N\sigma_j^x(a+a\yd)\;. \label{eq:Dicke}\ee
Here $\omega_0$ and $\omega_z$ are respectively the detuning of the cavity and of the atoms,  $g$ is the atom-cavity coupling,
$[a,a\yd]=1$, $\sigma^z_j=\pm1/2$, $[\sigma^x_j,\sigma^y_j]=i\sigma^z_j$. For simplicity we assumed that all the atoms are identical, although the present approach can be immediately generalized to the inhomogeneous case. 

The Hamiltonian (\ref{eq:Dicke}) commutes with the total spin operator
$S=(S^x)^2+(S^y)^2+(S^z)^2$, where $S^\alpha=\sum_{j=1}^N \sigma^\alpha_j$. Thanks to this symmetry it is possible to decouple the $2^N$ spin states into block-diagonal Dicke manifolds with a well defined total spin $S \lesssim N/2$. 
%
This analysis reveals that the equilibrium Dicke model presents a continuous phase transition between a normal and a superradiant phases, both at zero and finite temperatures \cite{hepp73,wang73,narducci1973energy,charmichael1973higher,duncan1974effect,hillery1985semiclassical,emary03}. The Dicke transition signals the spontaneous symmetry breaking of a descrete $Z_2$ symmetry ($\sigma^x\to-\sigma^x$ and $a\to-a$) and belongs to the mean-field universality class \cite{vidal2006finite,chen2008numerically,dalladiehl}.


Following the theoretical proposal of Refs.[\onlinecite{domokos2002collective,dimer07,nagy10}], the Dicke transition was recently realized in driven-dissipative quantum optical systems\cite{black03,baumann10,baumann2011exploring,brennecke2013real,baden2014realization,klinder2015dynamical,klinder2015observation}. The theoretical description of this transition\cite{garraway2011dicke,nagy11,oeztop11,bhaseen12}  considered the effect of the cavity decay $\kappa$, modeled as a Markovian bath coupled to the cavity field $a$.
This dissipative channel conserves the total spin and can be described through
a semiclassic Holstein-Primakoff\cite{holstein40,hillery1985semiclassical} approximation in which the total-spin operators are substituted by the bosonic operators $b$ and $b\yd$, according to $S^z\to - N/2 + b\yd b$ and $S^x \to \sqrt{N}(b+b\yd)$. 
This analysis leads to the critical coupling
\be g_c=\frac12\sqrt{\omega_z\frac{\omega^2_0+\kappa^2}{\omega_0}}\label{eq:meanfield}\;.
\ee
For $\kappa\to0$, Eq.~(\ref{eq:meanfield}) recovers the known equilibrium result. This  semiclassical approach relies on the conservation of the total spin and cannot be generalized to the case of single-atom dissipative processes.

%
%

{\bf Majorana fermions} To describe the atomic dephasing and decay we employ a fermionic path integral approach that allows us to expand the Dicke model in a $1/N$ series and resum all the leading terms \footnote{Path integrals offer a simple method to organize time-dependent perturbation theory. The same results can be alternatively obtained using for example the Nakajima–-Zwanzig approach (see Ref.[\onlinecite{breuer_book}] for an introduction).}. We specifically consider the Majorana-fermion  representation of spin-1/2 systems\cite{tsvelik2007quantum,shnirman2003spin,schad2015majorana}$^{,}$\footnote{Not to be confused with the Majorana representation of spins.}, $\sigma^z_j = f\yd_j f_j-1/2$, and $\sigma^+_j =\eta_j f_j\;$.
Here $f_j$ are Dirac fermions whose occupied (unoccupied) states correspond to spin-up (spin-down) states of the $j$-th atom and $\eta_j$ are Majorana fermion satisfying $\eta\yd_j=\eta_j$ and $\eta_j^2=1$. The role of these latter operators is essentially to map the commutation relations of the spins to the anticommutation relations of the fermions. Under this transformation the Dicke model (\ref{eq:Dicke}) becomes
\be H=\omega_0a\yd a - \omega_z\sum_{j=1}^N f\yd_j f_j + \frac{g}{\sqrt{N}}\sum_{j=1}^N \eta_j(f_j - f_j\yd)(a\yd+a)\;. \ee

Following the usual path-integral prescription we first introduce the bare Green functions describing the cavity and the fermions, and then derive Feynman rules for their coupling\footnote{See Ref.~[\onlinecite{sieberer2016keldysh}] for an introduction to Keldysh path-integrals in the context of quantum optics.}. In this study we focus on the {\it long-time} steady state in which all decay processes had time to stabilize and the Green functions depend on the time-difference only.
The bare (retarded) Green function of the cavity is then given in Ref.[\onlinecite{dalladiehl}] and equals to a $2\times2$ diagonal matrix, $G^{-1,R}_{a}(\omega)=(\omega+i\kappa)\tau_z - \omega_01_z$. Here $\tau_z$ is a Pauli matrix whose entries correspond to particles ($a\yd$) and holes ($a$) and $1_z$ is the unit matrix. The bare Green functions of the atoms describe their dynamics in the absence of the photon-atom coupling. We assume that each atom is coupled to an independent dissipative channel, leading to Green functions that do not couple different atoms and shall be denoted by $G_{f_j}(\omega)$ and $G_{\eta_j}(\omega)$.

\begin{figure}[b]
\includegraphics[scale=0.2]{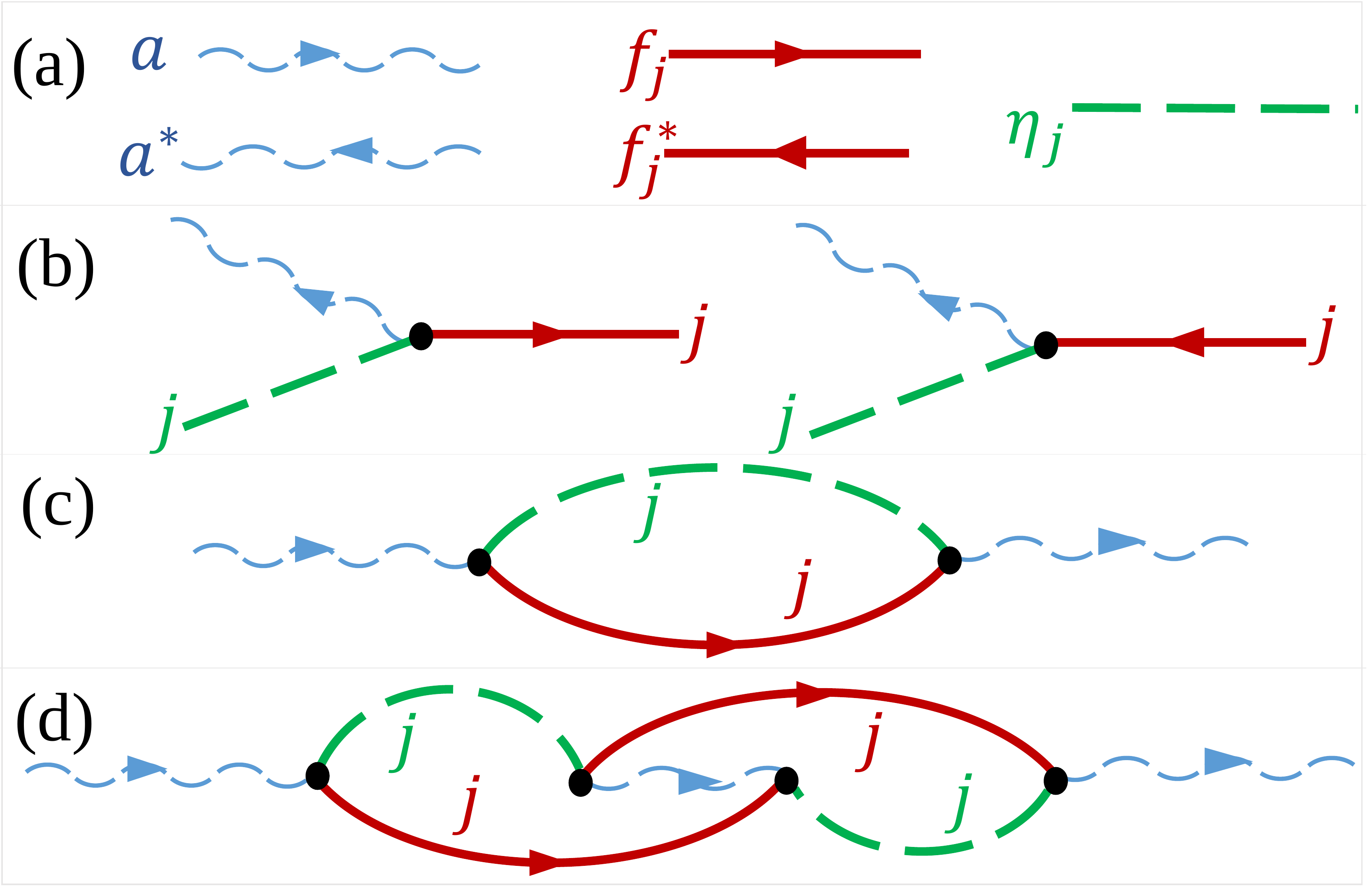}
\caption{(a) Bare Green functions. (b) Bare vertexes, proportional to $g/\sqrt{N}$. (c-d) One-loop and two-loop self energies for the cavity field $\Sigma^R_a$. The former contribution does not scale with $N$, while the latter scales as $1/N$ and can be neglected in the limit of $N\to\infty$.}
\label{fig:diagrams3}
\end{figure}

We next introduce the Rabi coupling as a vertex connecting the cavity field $a$, a fermionic field $f_j$, and a Majorana field $\eta_j$, with coefficient $g/\sqrt{N}$. This coupling generates a self-energy for the cavity field of the form $\Sigma_a^R(\omega)(1_z+\tau_z)$.
The Dicke transition corresponds to a diverging response function at $\omega=0$, or equivalently to a zero-frequency pole, and is set by
\begin{align}
{\rm det}\left[G_a^{-1,R}(0)+\Sigma_a^R(0)\right] = 0\end{align}
Substituting the expression for $G_a^{-1,R}$ we obtain
 \begin{align}
 &{\rm det}\left[\left(\ba{c c}i\kappa-\omega_0+\Sigma_a^R(0) & \Sigma_a^R(0) \\ \Sigma_a^R(0) & -i\kappa-\omega_0+\Sigma_a^R(0)\ea\right)\right] = 0\nn\;,
 \end{align}
 leading to the critical condition
 \begin{align} \omega^2_0+\kappa^2+2\omega_0\Sigma^R_a(0) = 0\label{eq:critical}
\end{align}
In general, the self energy $\Sigma_a^R$ depends on the photon-atom coupling $g$ and Eq.~(\ref{eq:critical}) sets its critical value, $g_c$. 



{\bf 1/N expansion} To compute the self-energy $\Sigma^R_a(\omega)$ we need to consider all possible diagrams that start and end with a cavity field (see  Fig.~\ref{fig:diagrams3} for details). A one-loop diagram is plotted in Fig.~\ref{fig:diagrams3}(c), and equals to\footnote{For simplicity here we assume that $G^K_f(\omega)$ is diagonal in Nambu space and that $G^K_\eta(\omega)=0$. The final expression, Eq.~(\ref{eq:sigmaR}), does not rely on these assumptions.}
\begin{align}
\Sigma^R_a(\omega) =\frac{g^2}{N}\sum_{j=1}^N~&\int_{-\infty}^\infty \frac{d\Omega}{2\pi}~\left[G^K_{f_j}(\Omega) G^R_{\eta_j}(\omega-\Omega)\right.
\nn\\
&+\left.G^K_{f_j}(-\Omega) G^R_{\eta_j}(\omega-\Omega)\right]\label{eq:sigmaGG}
\end{align}
Here the second term is generated by a diagram analogous to Fig.~\ref{fig:diagrams3}(c), but with an inverse direction of the fermionic arrow. Note that the resulting integral does not depend on $N$: each vertex introduces a $1/\sqrt{N}$ factor, balanced by the sum over all atoms. Fig.~\ref{fig:diagrams3}(d) shows an irreducible two-loop integral that contributes to the self energy of the cavity field. This diagram contains four vertexes and a single sum over $j$ and is therefore suppressed as $1/N$. (See also Refs.~[\onlinecite{piazza2013bose,piazza2014quantum}] for a similar result in the case of atoms with motional degrees of freedom.). In the limit of $N\to\infty$ only series of one-loop irreducible diagrams do not vanish. This series is exactly resummed by the above-mentioned self-energy approach. 

The self-energy (\ref{eq:sigmaGG}) has a simple interpretation in terms of spin-spin correlation functions. To see this mapping it is convenient to transform the integral expression appearing in Eq.~(\ref{eq:sigmaGG}) to the time domain
\begin{align}
\Sigma_a^R(\omega)=&\frac{{\rm i}g^2}{N}\sum_{j=1}^N \int_0^\infty dt \av{[f_j(t)\eta_j(t),f\yd_j(0)\eta_j(0)]}~e^{i\omega t}\nn\\
&~~~~~~~~~+\av{[f\yd_j(t)\eta_j(t),f_j(0)\eta_j(0)]}~e^{i\omega t}\\
=& \frac{4{\rm i}g^2}N \sum_{j=1}^N\int_0^\infty dt~ \av{[\sigma^x_j(t),\sigma^x_j(0)]}~e^{i\omega t}\\
=&-\frac{8g^2}N \sum_{j=1}^N\int_0^\infty dt~{\rm{Im}}\left[\av{\sigma^x_j(t)
\sigma^x_j(0)}\right]~e^{i\omega t}\;.
 \label{eq:sigmaR}
\end{align} 
Here the average $\av{...}$ refers to the bare theory in which the atoms are decoupled from the cavity, in analogy to the Lamb theory of the lasing transition \cite{lamb1964theory,agarwal1990steady,scully97,gartner2011two}: Eq.~(\ref{eq:sigmaR}) involves a sum over $j$, indicating that in the limit of $N\to\infty$, the cavity feels each atom independently.


Eqs.~(\ref{eq:critical}) and (\ref{eq:sigmaR}) express the position of the Dicke transition in terms of the correlation functions of individual dissipative spins. These correlations can be computed using either the Majorana fermion representation\cite{tsvelik2007quantum,shnirman2003spin,schad2015majorana}, or more conventional methods of quantum optics, such as master equations in the Lindblad form. For the sake of brevity, we employ here this latter method and leave the corresponding calculations using Majorana fermions for a future longer study. The introduction of Majorana fermions in the present work was nevertheless necessary to develop the $1/N$ expansion leading to Eq.~(\ref{eq:sigmaR}).


We specifically consider three distinct types of single-atom baths, listed in  Fig.~\ref{fig:schematic} along with their corresponding Lindblad operators:

{\bf (i) Dephasing --} Dephasing processes preserve the spin polarization of the atoms and can be mathematically described by the Lindblad operators $\sigma^z_j$. In the presence of this type of dissipation, the spin-spin correlation functions can be computed using the master equation: for any $t>0$ one finds (see Methods below) 
\be \av{\sigma^x_j(t)\sigma^x_j(0)} = e^{-\gamma_\phi t}\left[ \cos(\omega_z t) + i\av{\sigma_z} \sin(\omega_z t)\right]\label{eq:corr}\;.\ee
Combining this expression with Eqs.~(\ref{eq:critical}) and (\ref{eq:sigmaR}) we find
\be \Sigma_a^R(0) = 4g^2\frac{\av{\sigma^z_j}\omega_z}{\omega_z^2+\gamma_\phi^2}\;~~{\rm and~~} g_c =\frac12 \sqrt{\frac{\omega_z^2+\gamma_\phi^2}{-2\av{\sigma^z_j}\omega_z}\frac{\omega^2_0+\kappa^2}{\omega_0}}.\label{eq:gc_phi}\ee 

Since this specific type of bath preserves $\sigma^z_j$, its expectation value  is determined by the initial condition of the atoms. Importantly, if the initial state has $\av{\sigma^z_j}=0$ the Dicke transition does not occur ($g_c\to\infty$). As we will see below, the realization of a steady state with $\av{\sigma^z_j}\neq 0$ is actually a sufficient condition for the observation of the Dicke transition. As already observed by Refs.[\onlinecite{wolfe2014certifying,santos2016elucidating}] the spins do not need to form a coherent/entangled state to support this transition.

{\bf (ii) Thermal bath --} Let us now consider a decay channel induced by a thermal bath at temperature $T$, with decay rate $\gamma_T$. This situation is equivalent to having two Lindblad baths respectively coupled to $\sigma^-_j$ and $\sigma^+_j$ with rates $(1+n_T)\gamma_T$ and $n_T\gamma_T$, where $n_T$ is the Bose-Einstein distribution (see Methods section). Eq.~(\ref{eq:gc_phi}) is modified according to $\av{\sigma^z}\to 0.5~{\rm tanh}(\omega_z/2T)$ and $\gamma_\phi\to \gamma_T / {\rm tanh}(\omega_z/2T)$. The critical coupling is then given by
\be
g_c = \frac12\sqrt{\frac{\omega_z^2{\rm tanh}^2(\omega_z/2T)+\gamma_T ^2}{\omega_z {\rm tanh}^3(\omega_z/2T)}\frac{\omega_0^2+\kappa^2}{\omega_0}}\label{eq:gc_T}\;.
\ee
%
In the limit of $\gamma_T\to0$ and $\kappa\to0$, Eq.~(\ref{eq:gc_T}) reproduces the critical temperature of the equilibrium closed system\cite{hepp73,wang73,charmichael1973higher,duncan1974effect}, given by ${\rm tanh}(\omega_z/2T) = \omega_z\omega_0/4g_c^2$.

In general, Eq.(\ref{eq:gc_T}) is a monotonous increasing function of the temperature indicating that as expected, the superradiant transition is suppressed by the temperature of the spins. Interestingly, Eq.~(\ref{eq:gc_T}) shows that the critical temperature is affected by the decay rates $\kappa$ and $\gamma_T$. This result is in striking contrast to the common classical equilibrium case, where the strength of the coupling to a dissipative bath is not expected to affect the critical temperature\cite{hohenberg77}.

\begin{figure}[b]
\includegraphics[scale=0.8]{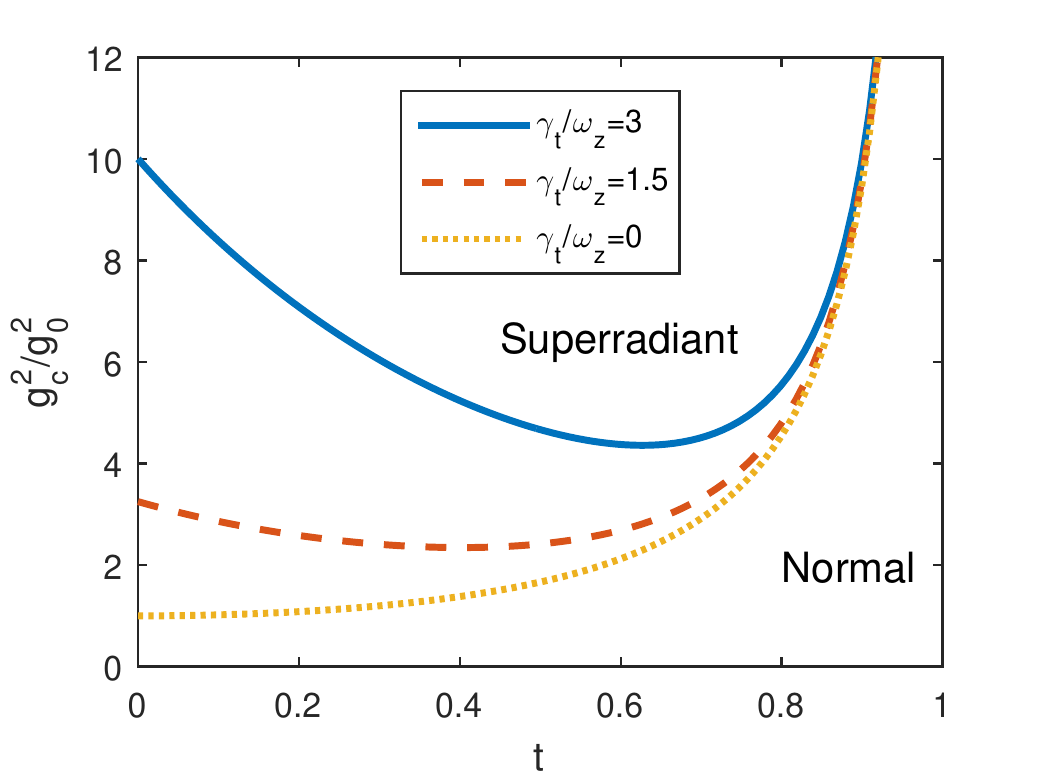}
\caption{Critical coupling of the Dicke model in the presence of a Markovian single-atom decay channel described by the Lindblad operator $L=\sigma^-+t\sigma^+$. Here $g_0$ is the critical coupling for a fully polarized system, Eq.~(\ref{eq:meanfield}). The critical coupling diverges in the limit of $t\to1$, where $\av{\sigma^z_j}=0$.}
\label{fig:theory}
\end{figure}

{\bf (iii) Generalized Markovian bath --} We finally consider a Markovian bath that couples coherently to both $\sigma_j^-$ and $\sigma_j^+$, and is described by the Lindblad operator $L_j=\sigma_j^- + t\sigma_j^+$, where $t$ is a fixed parameter. This situation might be relevant to some implementations of Dicke-type models using the 4-level scheme of Ref.~[\onlinecite{dimer07}] (see Ref.~[\onlinecite{dallaotter}] for details). A straightforward calculation (see Methods below) shows that the critical coupling is given by Eq.~(\ref{eq:gc_phi}) with $\av{\sigma^z}=0.5(1-t^2)/(1+t^2)$ and $\gamma_\phi \to \gamma_{\rm eff} = \gamma_t(1-t)^2$, leading to the critical coupling
\be
g_c =\frac12 \sqrt{\frac{(1+t^2)(\omega_z^2+\gamma_t^2(1-t)^2)}{(1-t^2)\omega_z}\frac{\omega^2_0+\kappa^2}{\omega_0}}\;.\label{eq:gc_t}
\ee
In the limit $t\to0$ we recover the semiclassic result of Ref.~[\onlinecite{strack2016}]: in this case the steady states coincides with the fully polarized state $\prod_j\ket{\downarrow_z}_j$ and the Holstein-Primakoff approximation becomes exact. In the opposite limit $t\to1$, the Dicke transition does not occur because the steady state is characterized by $\av{\sigma^z_j}=0$, in contrast to the result of the non-linear sigma model of Ref.~[\onlinecite{dalladiehl}].

For intermediate $0<t<1$, the interplay between $\gamma_{\rm eff}$ and $\av{\sigma^z}$ leads to the non-trivial behavior depicted in Fig.~\ref{fig:theory}. Note in particular that $\gamma_{\rm eff}$ is a decreasing function of $t$ and tends to 0 at $t=1$, in analogy to the spontaneous-emission-induced coherence of Ref.~[\onlinecite{gonzalez2013mesoscopic}]. Indeed in this limit the Lindblad operator is $\sigma_x$ and does not directly affect the correlator $\av{\sigma^x(t)\sigma^x(0)}$. As a consequence, for small $t\ll 1$, $g_c(t)$ has a negative slope due to the linear decrease of $\gamma_{\rm eff}$ as a function of $t$.  In contrast, for $t\lesssim 1$, $g_c(t)$ has a positive slope due to the decrease of $\av{\sigma^z} \sim (1-t^2)$. The resulting non-monotonous behavior differs from the previously-studied collective decay channels, where the critical coupling depends on the effective decay rate only.


{\bf Conclusion} 
In summary we studied the effects of atomic decay channels on the Dicke transition. Employing a fermionic path-integral analysis, we derived a closed expression for the critical coupling in terms of single-atom correlations, Eqs.~(\ref{eq:critical}) and (\ref{eq:sigmaR}). We considered several types of dissipative channels and computed the correspondent value of the critical photon-atom coupling $g_c$. We found that in general the critical coupling does not depend on the total spin of the system $S$, but rather on the average spin polarization $\av{\sigma^z_j}$, (see Eqs.~(\ref{eq:critical}) and (\ref{eq:gc_phi})). If the dissipative channel leads to a depolarized steady-state with $\av{\sigma^z_j}=0$ the Dicke transition disappears. 

In the present discussion we considered non-equilibrium steady states, in which all correlation functions depend on the time-difference only. The present analysis can nevertheless be directly extended to the study of the real-time dynamics, by considering the {\it retarded} Green function $\theta(t-t')\av{[\sigma^x_j(t),\sigma^x_j(t')]}$. In analogy to the steady state situation, it is sufficient to first solve for the dynamics of each atom independently, and then use this result to compute the response of the cavity. This approach might be useful to describe the transient Dicke transition observed in the experiments\cite{baden2014realization,klinder2015dynamical,klinder2015observation,roof2016observation}. 

It is also possible to extend the present analysis to realistic experimental situations including for example: the coexistance of thermal and Markovian baths; non-symmetric Dicke models where the rotating and counter-rotating terms of the Dicke Hamiltonian are different; multi-mode cavities where glassy transitions are expected\cite{strack2011dicke,gopalakrishnan2011frustration,buchhold2013dicke,mehrtash2015far}. Finally, it would be interesting to study the critical exponents of the non-equilibrium transition and compare them with the equilibrium case, following the lines of Ref.~[\onlinecite{dalladiehl}].

{\bf Acknowledgments} We acknowledge useful discussions with H. Tureci and P. Strack. This work is
supported by the Israel Science Foundation Grant No. 1542/14, Harvard-MIT CUA, NSF Grants No. DMR-1308435 and PHY-1506284, MURI-AFOSR, ARO-MURI Atomtronics, ARO-MURI Qusim, M. Rossler,
the Walter Haefner Foundation, the Humboldt Foundation, the Simons Foundation,
and the ETH Foundation.

\section*{METHODS: Master equations for a single spin coupled to a dissipative bath }

In this section we use the Lindblad master equation\cite{breuer_book} to compute the correlations of a single spin in the presence of dissipation. We then apply Eq.~(\ref{eq:sigmaR}) and compute the cavity self-energy. These calculations are not explicitly mentioned in the main article because they are completely standard, and are brought here only for the sake of completeness.

We consider an isolated spin described by the Hamiltonian $H=\omega_z\sigma_z$  and the Lindblad operator $L$. To compute $S_x(t)= \av{\sigma^x(t)\sigma^x(x)}$ we first derive the time evolution of the operator $\sigma^x(t)$ from the master equation
\be \frac{d\sigma^x}{dt}=-i[H,\sigma^x] -\gamma_\a\left(L_\a L_\a\yd \sigma^x + \sigma^x L_\a L_\a\yd -2 L\yd \sigma^x L\right)\label{eq:master}\ee

\def\subsection*#1{{\bf #1} --}

\subsection* {(i) Dephasing}
\label{app:A_phi}
For $L=\sigma^z$, Eq.~(\ref{eq:master}) becomes
\begin{align} 
\frac{d\sigma^x(t)}{dt}&=\omega_z\sigma_y   - {\gamma_\phi} \sigma_x\label{eq:eom1}\\
\frac{d\sigma^y(t)}{dt}&=-{\omega_z}\sigma_x   - {\gamma_\phi}\sigma_y\label{eq:eom2}
\end{align}
These equations are solved by
\begin{align} 
\sigma_x(t) &= e^{-\gamma_\phi t}\left(\cos(\omega_z t)\sigma_x(0) + \sin(\omega_z t)\sigma_y(0)\right) \\
\sigma_y(t) &= e^{-\gamma_\phi t}\left(\cos(\omega_z t)\sigma_y(0) - \sin(\omega_z t)\sigma_x(0)\right)
\end{align}
Using $(\sigma^x_j)^2=1/4$ we find that for any $t>0$  
\begin{align} S_x(t) &= \av{\sigma^x(t)\sigma^x(0)} 
\\&= \frac14 e^{-\gamma_\phi t}\left(\cos(\omega_z t) - 2i \av{\sigma_z} \sin(\omega_z t)\right)\;.\end{align}
 A straightforward integration gives:
\be \Sigma_a^R = 4g^2 \int_0^\infty S_x(t) - S_x(-t) = \frac{4g^2\omega_z \av{\sigma_z}}{(\omega_z^2+\gamma^2)} \label{eq:gammaz}
\ee

\subsection* { (ii) Thermal bath}
\label{app:A_T}
We now consider the decay process due to the coupling to a finite temperature bath. The correspondent master equation is
\be \frac{d\sigma^x}{dt}=-i[H,\sigma^x] -\sum_{\a=\pm}\gamma_\a\left(L_\a L_\a\yd \sigma^x + \sigma^x L_\a L_\a\yd -2 L\yd \sigma^x L\right)\label{eq:master_T}\;,\ee
where $\gamma_-=n_T + 1$, $\gamma_+=n_T$, and $n_T=(e^{\omega_z/T}-1)^{-1}$ is the Bose-Einstein distribution\cite{carmichael2009open}. A direct evaluation demonstrates that Eqs.~(\ref{eq:eom1}) and (\ref{eq:eom2}) are modified by $\gamma_\phi\to(1+2n_T)\gamma_{T} =\gamma_T/{\rm tanh}(\omega_z/2T)$. Using the corresponding master equation for $\sigma^z$ one finds
\begin{align}
\frac{d\sigma^z_j}{dt} &= \gamma_T(1+n_T) (2\sigma^z-1) + n_T (2\sigma^z+1) \\
&=\gamma_T\left[-1 + (2+4n_T)\sigma^z\right]
\end{align}
Thus, in the steady state $\av{\sigma^z}=1/(2+4n_T)=0.5~ {\rm tanh}(\omega_z/T)$, in agreement with the equilibrium result.

\subsection*{ (iii) Generalized Markovian bath}
\label{app:A_t}
We now consider the Lindblad operator $L=\sigma^-+t\sigma^+$. A direct evaluation leads to
\be L\yd L\sigma^x + \sigma^x L\yd L - 2 L\yd\sigma^x L  = \gamma_t(1-t)^2 \sigma^x\ee
As a consequence, the equations of motion of $\sigma^x$ are the same as (\ref{eq:eom1}) and (\ref{eq:eom2}), with $\gamma_\phi\to \gamma_t(1-t)^2$. Note in particular that for $t=1$, $L=\sigma^x$ and the correlator of $\sigma^x$ does not decay over time. We deduce that 
\be \Sigma^R_a(0)=\frac{4g^2\omega_z\av{\sigma^z}}{\omega_z^2+\gamma_t^2(1-t^2)}\label{eq:sigma_gamma_t}\ee
We next need to find the steady state expectation value of $\av{\sigma^z_j}$. For this purpose we use the master equation (\ref{eq:master}) with $\sigma^x\to\sigma^z$ and obtain
\be 
\frac{d\sigma^z_j}{dt} = -\gamma_t \left[(1-t^2) - 2(1+t^2)\sigma^z\right]
\ee
In the steady state the expectation value of the LHS is zero and $\av{\sigma^z} =0.5 (1-t^2)/(1+t^2)$. Combining this expression with Eq.~(\ref{eq:sigma_gamma_t}) we obtain Eq.~(\ref{eq:gc_t}).


\begin{thebibliography}{10}

\bibitem{takeuchi2007directed}
Takeuchi, K.~A., Kuroda, M., Chat\'e, H., and Sano, M.
\newblock {\em Phys. Rev. Lett.}{ \bf 99}, 234503 Dec  (2007).

\bibitem{henkel2008non}
Henkel, M., Hinrichsen, H., L{\"u}beck, S., and Pleimling, M.
\newblock {\em Non-equilibrium phase transitions}, volume~1.
\newblock Springer,  (2008).

\bibitem{black03}
Black, A.~T., Chan, H.~W., and Vuleti\ifmmode~\acute{c}\else \'{c}\fi{}, V.
\newblock {\em Phys. Rev. Lett.}{ \bf 91}, 203001 Nov  (2003).

\bibitem{baumann10}
Baumann, K., Guerlin, C., Brennecke, F., and Esslinger, T.
\newblock {\em Nature}{ \bf 464}, 1301--1306 (2010).

\bibitem{baumann2011exploring}
Baumann, K., Mottl, R., Brennecke, F., and Esslinger, T.
\newblock {\em Phys. Rev. Lett.}{ \bf 107}, 140402 Sep  (2011).

\bibitem{brennecke2013real}
Brennecke, F., Mottl, R., Baumann, K., Landig, R., Donner, T., and Esslinger,
  T.
\newblock {\em Proceedings of the National Academy of Sciences}{ \bf 110}(29),
  11763--11767 (2013).

\bibitem{baden2014realization}
Baden, M.~P., Arnold, K.~J., Grimsmo, A.~L., Parkins, S., and Barrett, M.~D.
\newblock {\em Phys. Rev. Lett.}{ \bf 113}, 020408 Jul  (2014).

\bibitem{klinder2015dynamical}
Klinder, J., Ke{\ss}ler, H., Wolke, M., Mathey, L., and Hemmerich, A.
\newblock {\em Proceedings of the National Academy of Sciences}{ \bf 112}(11),
  3290--3295 (2015).

\bibitem{klinder2015observation}
Klinder, J., Ke\ss{}ler, H., Bakhtiari, M.~R., Thorwart, M., and Hemmerich, A.
\newblock {\em Phys. Rev. Lett.}{ \bf 115}, 230403 Dec  (2015).

\bibitem{hepp73}
Hepp, K. and Lieb, E.~H.
\newblock {\em Annals of Physics}{ \bf 76}(2), 360 -- 404 (1973).

\bibitem{wang73}
Wang, Y.~K. and Hioe, F.~T.
\newblock {\em Phys. Rev. A}{ \bf 7}, 831--836 Mar  (1973).

\bibitem{narducci1973energy}
Narducci, L.~M., Orszag, M., and Tuft, R.~A.
\newblock {\em Phys. Rev. A}{ \bf 8}, 1892--1906 Oct  (1973).

\bibitem{charmichael1973higher}
Carmichael, H., Gardiner, C., and Walls, D.
\newblock {\em Physics Letters A}{ \bf 46}(1), 47 -- 48 (1973).

\bibitem{duncan1974effect}
Duncan, G.~C.
\newblock {\em Physical Review A}{ \bf 9}(1), 418 (1974).

\bibitem{hillery1985semiclassical}
Hillery, M. and Mlodinow, L.~D.
\newblock {\em Phys. Rev. A}{ \bf 31}, 797--806 Feb  (1985).

\bibitem{emary03}
Emary, C. and Brandes, T.
\newblock {\em Phys. Rev. E}{ \bf 67}, 066203 Jun  (2003).

\bibitem{dalladiehl}
Dalla~Torre, E.~G., Diehl, S., Lukin, M.~D., Sachdev, S., and Strack, P.
\newblock {\em Phys. Rev. A}{ \bf 87}, 023831 Feb  (2013).

\bibitem{strack2016}
Gelhausen, J., Buchhold, M., and Strack, P.
\newblock {\em arXiv preprint arXiv:1605.07637}{ \bf } (2016).

\bibitem{vidal2006finite}
Vidal, J. and Dusuel, S.
\newblock {\em EPL (Europhysics Letters)}{ \bf 74}(5), 817 (2006).

\bibitem{chen2008numerically}
Chen, Q.-H., Zhang, Y.-Y., Liu, T., and Wang, K.-L.
\newblock {\em Phys. Rev. A}{ \bf 78}, 051801 Nov  (2008).

\bibitem{domokos2002collective}
Domokos, P. and Ritsch, H.
\newblock {\em Phys. Rev. Lett.}{ \bf 89}, 253003 Dec  (2002).

\bibitem{dimer07}
Dimer, F., Estienne, B., Parkins, A.~S., and Carmichael, H.~J.
\newblock {\em Phys. Rev. A}{ \bf 75}, 013804 Jan  (2007).

\bibitem{nagy10}
Nagy, D., K\'onya, G., Szirmai, G., and Domokos, P.
\newblock {\em Phys. Rev. Lett.}{ \bf 104}, 130401 Apr  (2010).

\bibitem{garraway2011dicke}
Garraway, B.~M.
\newblock {\em Philosophical Transactions of the Royal Society of London A:
  Mathematical, Physical and Engineering Sciences}{ \bf 369}(1939), 1137--1155
  (2011).

\bibitem{nagy11}
Nagy, D., Szirmai, G., and Domokos, P.
\newblock {\em Phys. Rev. A}{ \bf 84}, 043637 Oct  (2011).

\bibitem{oeztop11}
\"Oztop, B., Bordyuh, M., M\"ustecaplio\u{g}lu, O.~E., and T\"{u}reci, H.~E.
\newblock {\em New Journal of Physics}{ \bf 14}(8), 085011 (2012).

\bibitem{bhaseen12}
Bhaseen, M.~J., Mayoh, J., Simons, B.~D., and Keeling, J.
\newblock {\em Phys. Rev. A}{ \bf 85}, 013817 Jan  (2012).

\bibitem{holstein40}
Holstein, T. and Primakoff, H.
\newblock {\em Phys. Rev.}{ \bf 58}, 1098--1113 (1940).

\bibitem{tsvelik2007quantum}
Tsvelik, A.~M.
\newblock {\em Quantum field theory in condensed matter physics}.
\newblock Cambridge university press,  (2007).

\bibitem{shnirman2003spin}
Shnirman, A. and Makhlin, Y.
\newblock {\em Physical review letters}{ \bf 91}(20), 207204 (2003).

\bibitem{schad2015majorana}
Schad, P., Makhlin, Y., Narozhny, B., Sch{\"o}n, G., and Shnirman, A.
\newblock {\em Annals of Physics}{ \bf 361}, 401--422 (2015).

\bibitem{piazza2013bose}
Piazza, F., Strack, P., and Zwerger, W.
\newblock {\em Annals of Physics}{ \bf 339}, 135--159 (2013).

\bibitem{piazza2014quantum}
Piazza, F. and Strack, P.
\newblock {\em Phys. Rev. A}{ \bf 90}, 043823 Oct  (2014).

\bibitem{lamb1964theory}
Lamb~Jr, W.~E.
\newblock {\em Physical Review}{ \bf 134}(6A), A1429 (1964).

\bibitem{agarwal1990steady}
Agarwal, G. and Gupta, S.~D.
\newblock {\em Physical Review A}{ \bf 42}(3), 1737 (1990).

\bibitem{scully97}
Scully, M.~O. and Zubairy, M.~S.
\newblock {\em Quantum Optics}.
\newblock Cambridge University Press,  (1997).

\bibitem{gartner2011two}
Gartner, P.
\newblock {\em Phys. Rev. A}{ \bf 84}, 053804 Nov  (2011).

\bibitem{wolfe2014certifying}
Wolfe, E. and Yelin, S.
\newblock {\em Physical review letters}{ \bf 112}(14), 140402 (2014).

\bibitem{santos2016elucidating}
Santos, E. M.~d. and Duzzioni, E.~I.
\newblock {\em arXiv preprint arXiv:1604.08184}{ \bf } (2016).

\bibitem{hohenberg77}
Hohenberg, P.~C. and Halperin, B.~I.
\newblock {\em Rev. Mod. Phys.}{ \bf 49}, 435--479 Jul  (1977).

\bibitem{dallaotter}
Dalla~Torre, E.~G., Otterbach, J., Demler, E., Vuletic, V., and Lukin, M.~D.
\newblock {\em Phys. Rev. Lett.}{ \bf 110}, 120402 Mar  (2013).

\bibitem{gonzalez2013mesoscopic}
Gonz{\'a}lez-Tudela, A. and Porras, D.
\newblock {\em Physical review letters}{ \bf 110}(8), 080502 (2013).

\bibitem{roof2016observation}
Roof, S., Kemp, K., Havey, M., and Sokolov, I.
\newblock {\em arXiv preprint arXiv:1603.07268}{ \bf } (2016).

\bibitem{strack2011dicke}
Strack, P. and Sachdev, S.
\newblock {\em Physical review letters}{ \bf 107}(27), 277202 (2011).

\bibitem{gopalakrishnan2011frustration}
Gopalakrishnan, S., Lev, B.~L., and Goldbart, P.~M.
\newblock {\em Physical review letters}{ \bf 107}(27), 277201 (2011).

\bibitem{buchhold2013dicke}
Buchhold, M., Strack, P., Sachdev, S., and Diehl, S.
\newblock {\em Phys. Rev. A}{ \bf 87}, 063622 Jun  (2013).

\bibitem{mehrtash2015far}
Babadi, M., Demler, E., and Knap, M.
\newblock {\em Phys. Rev. X}{ \bf 5}, 041005 Oct  (2015).

\bibitem{breuer_book}
Breuer, H.-P. and Petruccione, F.
\newblock {\em The theory of open quantum systems}.
\newblock Oxford University Press on Demand,  (2002).

\bibitem{carmichael2009open}
Carmichael, H.
\newblock {\em An open systems approach to quantum optics: lectures presented
  at the Universit{\'e} Libre de Bruxelles, October 28 to November 4, 1991},
  volume~18.
\newblock Springer Science \& Business Media,  (2009).

\bibitem{sieberer2016keldysh}
Sieberer, L.~M., Buchhold, M., and Diehl, S.
\newblock {\em Reports on Progress in Physics}{ \bf 79}(9), 096001 (2016).

\end{thebibliography}

\vspace{0.01cm}

\end{document}